\documentstyle[preprint,revtex]{aps}
\tightenlines
\begin{document}
\draft

\preprint{SLAC--PUB--6059}
\medskip
\preprint{February 1993}
\medskip
\preprint{T/E}

\begin{title}
Reparametrization invariance and the expansion of currents\\
in the heavy quark effective theory
\end{title}

\author{Matthias Neubert\thanks{Work supported by the Department of
Energy under contract DE-AC03-76SF00515.}}
\begin{instit}
Stanford Linear Accelerator Center\\
Stanford University, Stanford, California 94309
\end{instit}

\begin{abstract}
The coefficients appearing at leading and subleading order in the $1/m$
expansion of bilinear heavy quark currents are related to each other by
imposing reparametrization invariance on both the effective current
operators and the short-distance coefficient functions in the heavy
quark effective theory. When combined with present knowledge about the
leading order coefficients, the results allow to calculate all
coefficients appearing at order $1/m$ to next-to-leading order in
renormalization-group improved perturbation theory. They also provide a
meaningful definition of the velocity transfer variable $v\cdot v'$ to
order $1/m$.
\end{abstract}
\centerline{(submitted to Physics Letters B)}
\newpage

\narrowtext

\section{Introduction}

Over the last few years, heavy quark effective field theory has been
established as an efficient tool to analyze decay processes involving
hadrons containing a heavy quark \cite{Eich,Geor,Mann,Falk,FGL}. In
such systems the heavy quark is almost on-shell and interacts with the
surrounding soup of light quarks, antiquarks and gluons predominantly
via the exchange of soft gluons. As $m_Q\gg\Lambda_{\rm QCD}$ these
soft interactions cannot resolve the structure of the heavy quark; in
particular, they are blind to its flavor and spin. In this limit the
heavy quark acts as a featureless color source. This is the origin of a
spin-flavor symmetry, which relates the properties of hadrons
containing different heavy quarks \cite{Volo,Isgu}.

Since, in the $m_Q\to\infty$ limit, the heavy quark velocity $v$ is
conserved with respect to soft QCD interactions, it is appropriate to
split the total momentum into a ``large'' kinetic piece and a ``small''
residual momentum $k$, which puts the heavy quark slightly off-shell:
$p_Q = m_Q v + k$. Because all dynamics resides in $k$, it is useful to
absorb the mass-dependent piece of the momentum by a field
redefinition. To this end, one introduces a velocity-dependent field by
\cite{Geor}
\begin{equation}\label{redef}
   Q(x) = e^{i m_Q v\cdot x}\,\,h_v(x)
\end{equation}
and imposes the on-shell condition $\rlap/v\,h_ v(x)=h_v(x)$,
corrections to which are suppressed as $\Lambda_{\rm QCD}/m_Q$. The new
fields carry the residual momentum $k$, which by construction does not
scale with $m_Q$. Their strong interactions are described by the
so-called heavy quark effective theory (HQET), which essentially
provides an expansion in $k/m_Q$. To lowest order in $1/m_Q$, the
effective Lagrangian is \cite{Geor,Mann}
\begin{equation}\label{Leff}
   {\cal{L}}_v = \bar h_v\,i v\!\cdot\!D\,h_v \,,
\end{equation}
where $D$ is the gauge-covariant derivative. Such a Lagrangian has to
be written for every heavy quark in the process under consideration.
For $N_h$ heavy quarks of the same velocity, the total Lagrangian is
then invariant under a $SU(2 N_h)$ spin-flavor symmetry group. This
symmetry is explicitly broken at order $1/m_Q$ by the presence of
higher dimension operators \cite{Eich,FGL}.

For the effective theory to provide a converging expansion it is
necessary that $k$ be of order $\Lambda_{\rm QCD}$. This implies that
the heavy quark velocity $v$ must be close to the velocity $v_h$ of the
hadron containing the heavy quark:
\begin{equation}
   v = v_h + {\cal{O}}\Big(\Lambda_{\rm QCD}/m_Q\Big) \,.
\end{equation}
This still allows some freedom in the choice of $v$, however. Instead
of using $(v,k)$ as the heavy quark velocity and residual momentum, one
can as well construct HQET using some different set of variables
$(v+q/m_Q,k-q)$, as long as $q$ is of order $\Lambda_{\rm QCD}$ and
satisfies $2 v\!\cdot\!q + q^2/m_Q=0$, so that the new velocity is
still a unit four-vector. The effective theories obtained in these two
ways must, of course, be equivalent \cite{Duga}. This so-called
reparametrization invariance of HQET is a very useful concept in that
it relates the coefficients of operators appearing at different order
in the $1/m_Q$ expansion \cite{RPI}. These relations are
renormalization-group invariant, i.e, they are true to all orders in
perturbation theory and cannot be subject to nonperturbative
corrections either.

In this paper, we use reparametrization invariance to derive relations
between the coefficients appearing at leading and subleading order in
the expansion of heavy quark currents in HQET. Some of these relations
were already obtained in Ref.~\cite{RPI}, by imposing reparametrization
invariance on the operators in the effective theory. However, it was
not realized until now that additional relations can be derived by
writing also the velocity-dependent short-distance coefficients in a
reparametrization invariant form. In fact, we show that {\it all\/}
coefficients of the effective current operators of dimension four can
be determined that way. In Sect.~2 we briefly discuss the heavy quark
expansion of currents in HQET. In Sect.~3 we recall the concept of
reparametrization invariance and study its implications for the
coefficients in the expansion of currents. Sect.~4 deals with an
interpretation of the reparametrization invariant extension of the
velocity transfer variable $v\cdot v'$. A short summary of the results
is given in Sect.~5.

\section{Expansion of currents in HQET}

Let us consider currents of the form $\bar Q'\,\Gamma\,Q$, which
mediate transitions between two heavy quarks $Q$ and $Q'$ of, in
general, different flavor. In principle $\Gamma$ could be an arbitrary
combination of Dirac matrices. For the weak currents, however,
$\Gamma=\gamma^\mu(1-\gamma_5)$. We are interested in hadronic matrix
elements of these currents between hadron states $H(v_h)$ and
$H'(v'_h)$ which contain the heavy quarks. HQET can be used to make the
dependence of such matrix elements on the heavy quark masses explicit.
In the effective theory each current has a representation as a series
of operators built from the new fields $h_v$ and $h'_{v'}$ replacing
$Q$ and $Q'$. These effective current operators can have dimension
higher than three, in which case they are multiplied by inverse powers
of the heavy quark masses. In general, one has
\begin{equation}\label{Jexp}
   \bar Q'\,\Gamma\,Q \,\widehat{=}\, \sum_i C_i\,J_i
   + \sum_j \Bigg[ {B_j\over 2 m_Q} + {B'_j\over 2 m_{Q'}} \Bigg] O_j
   + {\cal{O}}\big(1/m^2\big) \,,
\end{equation}
where the symbol $\widehat{=}$ is used for equations which are true in
matrix elements only, and $m$ stands generically for $m_Q$ or $m_{Q'}$.
Both the coefficients and the operators in this expansion depend on
$\Gamma$. The $\{J_i\}$ are a complete set of dimension three operators
with the same quantum numbers as the original current. Similarly, the
$\{O_j\}$ form a basis of dimension four operators. Since in the
effective theory the fields carry velocity labels, the effective
current operators can depend on $v$ and $v'$. In case of the vector
current $\bar Q'\,\gamma^\mu\,Q$, for instance, the dimension three
operators are
\begin{eqnarray}
   J_1 &=& \bar h'_{v'}\,\gamma^\mu\,h_v \,, \nonumber\\
   J_2 &=& \bar h'_{v'}\,v^\mu\,h_v \,, \nonumber\\
   J_3 &=& \bar h'_{v'}\,v'^\mu\,h_v \,,
\end{eqnarray}
while a convenient basis for the dimension four operators is
\begin{equation}\label{Odef}
 \begin{array}{ll}
   O_1 = \bar h'_{v'}\,\gamma^\mu\,i\,\rlap/\!D\,h_v \,, &\qquad
   O_8 = - \bar h'_{v'}\,i\,\overleftarrow{\rlap/\!D}\,\gamma^\mu\,h_v
   \,, \\
   O_2 = \bar h'_{v'}\,v^\mu\,i\,\rlap/\!D\,h_v \,, &\qquad
   O_9 = - \bar h'_{v'}\,i\,\overleftarrow{\rlap/\!D}\,v^\mu\,h_v
   \,, \\
   O_3 = \bar h'_{v'}\,v'^\mu\,i\,\rlap/\!D\,h_v \,, &\qquad
   O_{10} = - \bar h'_{v'}\,i\,\overleftarrow{\rlap/\!D}\,v'^\mu\,h_v
   \,, \\
   O_4 = \bar h'_{v'}\,i D^\mu\, h_v \,, &\qquad
   O_{11} = - \bar h'_{v'}\,i\overleftarrow{D^\mu}\,h_v
   \,, \\
   O_5 = \bar h'_{v'}\,\gamma^\mu\,i v'\!\cdot\!D\,h_v \,, &\qquad
   O_{12} = - \bar h'_{v'}\,i v\!\cdot\!\overleftarrow{D}\,
   \gamma^\mu\,h_v \,, \\
   O_6 = \bar h'_{v'}\,v^\mu\,i v'\!\cdot\!D\,h_v \,, &\qquad
   O_{13} = - \bar h'_{v'}\,i v\!\cdot\!\overleftarrow{D}\,v^\mu\,h_v
   \,, \\
   O_7 = \bar h'_{v'}\,v'^\mu\,i v'\!\cdot\!D\,h_v \,, &\qquad
   O_{14} = - \bar h'_{v'}\,i v\!\cdot\!\overleftarrow{D}\,v'^\mu\,h_v
    \,.
 \end{array}
\end{equation}
Similar sets of operators can be constructed for the expansion of the
axial vector current $\bar Q'\,\gamma^\mu\gamma_5\,Q$. In (\ref{Odef})
we have not included operators that vanish by the equation of motion $i
v\!\cdot\!D\,h_v=0$ following from the effective Lagrangian
(\ref{Leff}). They are irrelevant at the level of matrix elements. For
simplicity we have evaluated the currents at $x=0$; otherwise the
operators in HQET would acquire a phase according to (\ref{redef}).

Eq.~(\ref{Jexp}) provides a separation of short- and long-distance
contributions to current matrix elements. The perturbative corrections
arising from hard gluons (with virtualities of order $m_Q$ or $m_{Q'}$)
are factorized into the coefficients $C_i$, $B_j$ and $B'_j$, which are
functions of the heavy quark masses, the velocity transfer $v\cdot v'$,
as well as an arbitrary matching scale $\mu$: $C_i =
C_i(m_Q,m_{Q'},v\cdot v',\mu)$ etc. In particular, these functions
contain any logarithmic dependence on the heavy quark masses resulting
from the running couplings $\alpha_s(m_Q)$ and $\alpha_s(m_{Q'})$. All
long-distance effects, on the other hand, are still contained in the
hadronic matrix elements of the effective current operators, which are
to be evaluated between states of the effective theory. These matrix
elements can be parameterized by universal functions of the hadron
velocity transfer $v_h\!\cdot\!v'_h$, which are independent of the
heavy quark masses. They do depend on the matching scale, however, in
such a way that the right-hand side of (\ref{Jexp}) is
$\mu$-independent. For the vector and axial vector currents the
coefficients $C_i$ are known to next-to-leading order in
renormalization-group improved perturbation theory \cite{QCD}. The
coefficients $B_j$ and $B'_j$, on the other hand, have so far only been
calculated in leading logarithmic approximation \cite{AMM}.

\section{Relations imposed by reparametrization invariance}

The effective theory must be invariant under reparametrizations of the
heavy quark velocity and residual momentum which leave the total
momentum $p_Q = m_Q v + k$ unchanged. Luke and Manohar have
investigated the implications following from this simple statement in
detail \cite{RPI}. They found that the velocity and the covariant
derivative must always appear in the combination
\begin{equation}\label{Vdef}
   {\cal{V}} = v + {i D\over m_Q} \,,
\end{equation}
which can be interpreted as the gauge-covariant extension of the
operator $\widehat{p}_Q/m_Q$. A subtlety which has to be taken into
account is that the heavy quark spinor fields transform under
reparametrizations in a nontrivial way. They become invariant by
including a Lorentz boost $\Lambda({\cal{V}},v)$, which transforms $v$
into ${\cal{V}}$. The result is that the effective Lagrangian of HQET,
as well as any composite operator in the effective theory, must be
built of ${\cal{V}}$ and $\widetilde{h}_v = \Lambda({\cal{V}},v)\,h_v$.
At order $1/m$, the explicit form of $\widetilde{h}_v$ is
\begin{equation}\label{htdef}
   \widetilde{h}_v = \bigg(1 + {i\,\rlap/\!D\over 2 m_Q}\bigg)\,h_v
   = {1+\rlap/{\cal{V}}\over 2}\,h_v \,.
\end{equation}
Given this result, one can immediately relate some of the coefficients
in (\ref{Jexp}), namely \cite{RPI}:
\begin{eqnarray}\label{Brel1}
   B_1 &=& B'_8 = C_1 \,, \nonumber\\
   B_2 &=& {1\over 2}\,B_4 = B'_9 = C_2  \,, \nonumber\\
   B_3 &=& B'_{10} = {1\over 2}\,B'_{11} = C_3  \,.
\end{eqnarray}

Since all dimension four operators in (\ref{Odef}) contain a covariant
derivative acting on one of the heavy quark fields and are therefore
not reparametrization invariant by themselves, it is clear that there
must be additional relations. For instance, derivatives acting on $h_v$
can only come in combination with a coefficient $1/m_Q$, while those
acting on $\bar h_{v'}$ must come with $1/m_{Q'}$. Hence
\begin{eqnarray}
   B_j &= 0 \,; \quad &j=8,\ldots,14, \nonumber\\
   B'_j &= 0 \,; \quad &j=1,\ldots,7.
\end{eqnarray}
What remains to be determined, then, are the coefficients $B_j$ for
$j=5,6,7$ and $B'_j$ for $j=12,13,14$. The important new observation
which accomplishes this is that not only the effective current
operators, but also the {\it velocity-dependent coefficient
functions\/} must be written in a reparametrization invariant way. This
means that the variable $w=v\cdot v'$ which these functions depend on
has to be replaced by the reparametrization invariant {\it operator\/}
\begin{equation}
   \widehat{w} = {\cal{V}}'^\dagger\!\cdot{\cal{V}}
   = \bigg(v'-{i\overleftarrow{D}\over m_{Q'}}\bigg) \cdot
   \bigg(v+{iD\over m_Q}\bigg) \,,
\end{equation}
where it is understood that $iD$ acts only on $h_v$, while
$i\overleftarrow{D}$ acts on $\bar h'_{v'}$. Inserting the expansion
(for simplicity, we suppress the dependence of $C_i$ on the heavy quark
masses and on $\mu$)
\begin{equation}\label{Cexpand}
   C_i(\widehat{w}) = C_i(w) + {\partial C_i(w)\over\partial w}\,
   \Bigg[ {i v'\!\cdot\!D\over m_Q}
        - {i v\!\cdot\!\overleftarrow{D}\over m_{Q'}} \Bigg]
   + {\cal{O}}\big(1/m^2\big)
\end{equation}
into (\ref{Jexp}) one does indeed generate the remaining operators
$O_5$ to $O_7$ and $O_{12}$ to $O_{14}$. We find
\begin{eqnarray}\label{Brel2}
   B_5 &=& B'_{12} = 2\,{\partial C_1\over\partial w} \,, \nonumber\\
   B_6 &=& B'_{13} = 2\,{\partial C_2\over\partial w} \,, \nonumber\\
   B_7 &=& B'_{14} = 2\,{\partial C_3\over\partial w} \,.
\end{eqnarray}

Eqs.~(\ref{Brel1})--(\ref{Brel2}) summarize our main result:
Reparametrization invariance relates {\it all\/} coefficients appearing
at order $1/m$ in the heavy quark expansion of the vector current to
the coefficients appearing at leading order, and to their derivatives
with respect to $w=v\cdot v'$. A similar statement applies, of course,
for any other current. These relations are valid to all orders in
perturbation theory, and they cannot be modified by nonperturbative
corrections either.

At this point it is worthwhile to compare our exact results to some
approximate expressions for the short-distance coefficients known so
far in the literature. In Ref.~\cite{AMM} the coefficients have been
calculated in leading logarithmic approximation, working with an
average heavy quark mass $m$. In accordance with the relations
(\ref{Brel1}) and (\ref{Brel2}), one then obtains
\begin{eqnarray}
   C_1(w) &=& B_1(w) = B'_8(w)
    = \Bigg({\alpha_s(m)\over\alpha_s(\mu)}\Bigg)^{a_L(w)} \,,
    \nonumber\\
   B_5(w) &=& B'_{12}(w) = 2\,{\partial C_1(w)\over\partial w}
    = 2\,a_L'(w)\,\ln\!\Bigg({\alpha_s(m)\over\alpha_s(\mu)}\Bigg)\,
    C_1(w) \,,
\end{eqnarray}
where
\begin{equation}
   a_L(w) = {8\over 33-2 n_f}\,\Bigg[
   {w\over\sqrt{w^2-1}}\,\ln\big(w + \sqrt{w^2-1}\big) - 1 \Bigg] \,,
\end{equation}
and $n_f$ is the number of light quark flavors. All other coefficients
vanish in this approximation.\footnote{We can also compare to
Ref.~\cite{Cho}, where the matching contributions of order
$\alpha_s(m_{Q'})/m_{Q'}$ arising at $\mu=m_{Q'}$ have been computed.
The results given there satisfy the relations (\ref{Brel1}) and
(\ref{Brel2}). The expression presented for the coefficient $C_1$ is
incorrect, however. The correct result is given in Ref.~\cite{QCD}.}
Given our exact relations and the fact that the coefficients $C_i$ are
know to next-to-leading logarithmic order \cite{QCD}, it is now
possible to derive much more accurate expressions for $B_j$ and $B'_j$.

\section{Reparametrization invariant velocity transfer}

At order $1/m$, the effect of the operator $\widehat{w}$ in the
short-distance coefficients can be readily evaluated at the level of
matrix elements. The equation of motion $i v\cdot\!D\,h_v=0$ and the
corresponding equation for $\bar h'_{v'}$ allow one to replace the
covariant derivatives in (\ref{Cexpand}) by total derivatives acting on
the current, e.g.
\begin{equation}
   \bar h'_{v'}\,\Gamma\,i v'\!\cdot\!D\,h_v = i v'\!\cdot\!\partial\,
   \Big[ \bar h'_{v'}\,\Gamma\,h_v \Big] \,,
\end{equation}
where $\Gamma$ is again arbitrary. From translational invariance, and
taking into account the phase factors in the definition of the
effective heavy quark fields in (\ref{redef}), one finds that the
$x$-dependence of a current matrix element between hadron states
$H(v_h)$ and $H'(v'_h)$ is given by $\exp(-i\phi\cdot x)$, where
\begin{equation}
   \phi = (m_H\,v_h - m_Q\,v) - (m_{H'}\,v'_h - m_{Q'}\,v') \,.
\end{equation}
Using this, together with the fact that $m_H-m_Q = m_{H'}-m_{Q'}$ to
leading order in the $1/m$ expansion, it is straightforward to show that
\begin{equation}
   (\widehat{w}-1)\,\langle H'(v'_h)| \bar h'_{v'}\,\Gamma\,h_v
   |H(v_h)\rangle = {m_H m_{H'}\over m_Q m_{Q'}}\,(w_h - 1)\,
   \langle H'(v'_h)| \bar h'_{v'}\,\Gamma\,h_v |H(v_h)\rangle
   + {\cal{O}}(1/m^2) \,,
\end{equation}
where $w_h = v_h\!\cdot\!v'_h$. Note that the {\it hadron velocities\/}
appear in this equation. To order $1/m$, it follows that in matrix
elements the operator $\widehat{w}$ in the short-distance coefficient
functions $C_i$ can be replaced by the reparametrization invariant
velocity transfer variable
\begin{equation}\label{wbar}
   \bar w = 1 + {m_H m_{H'}\over m_Q m_{Q'}}\,(v_h\!\cdot\!v'_h-1) \,.
\end{equation}
If this variable is used in the coefficient functions, the operators
$O_5$ to $O_7$ and $O_{12}$ to $O_{14}$ no longer appear in the
expansion (\ref{Jexp}) since, for instance,
\begin{equation}
   C_1(w)\,J_1 + 2\,{\partial C_1(w)\over\partial w}\,
   \Bigg[ {O_5\over 2 m_Q} + {O_{12}\over 2 m_{Q'}} \Bigg]
   \,\widehat{=}\, C_1(\bar w)\,J_1 + {\cal{O}}\big(1/m^2\big) \,.
\end{equation}

Let us explore in more detail the physical meaning of the variable
$\bar w$. One might have expected that the reparametrization invariant
generalization of the quark velocity transfer would be the velocity
transfer of the hadrons, $w_h=v_h\!\cdot v'_h$. This is not the case,
however. Rather, in (\ref{wbar}) there appears an additional scaling
factor depending on the hadron and quark masses. The kinematic region
for $\bar w$ extends from $\bar w =1$ at zero recoil ($v_h\!\cdot\!v'_h
=1$) up to a maximum value given by
\begin{equation}
   \bar w_{\rm max} - 1 = {m_H m_{H'}\over m_Q m_{Q'}}\,
   {(m_H - m_{H'})^2\over 2 m_H m_{H'}}
   = {(m_Q - m_{Q'})^2\over 2 m_Q m_{Q'}} + {\cal{O}}(1/m^2) \,.
\end{equation}
This is just the maximum velocity transfer attainable in a decay of
{\it free\/} quarks. In fact, it can be readily seen that (\ref{wbar})
is precisely (up to terms of order $1/m^2$) the condition
\begin{equation}
   (p_H-p_{H'})^2 = (p_Q-p_{Q'})^2
\end{equation}
that the momentum transfer to the hadrons equals the momentum transfer
to free heavy quarks.

It is not hard to see why, away from zero recoil, the quark velocity
transfer $\bar w$ seen by hard gluons is different from the hadron
velocity transfer $w_h$. Consider the weak decay $H\to H'+W$. In the
initial state, the heavy quark $Q$ moves on average with the hadron's
velocity $v_h$. When the $W$ boson is emitted, the outgoing heavy quark
$Q'$ has in general some different velocity $v_{Q'}$. Over short time
scales this velocity remains unchanged, and this is what is seen by
hard gluons. After the $W$ emission, however, the light degrees of
freedom in the initial hadron still have the initial hadron's velocity.
They have to combine with the outgoing heavy quark to form the final
state hadron $H'$. This rearrangement happens over much larger,
hadronic time scales by the exchange of soft gluons. In this process
the velocity of $Q'$ is changed by an amount of order $1/m$ (its
momentum is changed by an amount of order $\Lambda_{\rm QCD}$). Hence
the hadron velocity transfer differs from the ``short-distance'' quark
velocity transfer by an amount of order $1/m$. The precise relation
between $w_h$ and $\bar w$ is determined by momentum conservation and
is given in (\ref{wbar}). At zero recoil, no such rearrangement is
needed, and indeed $\bar w=w_h=1$ in this limit.

\section{Summary}

We have shown that, to order $1/m$ in the heavy quark effective theory,
the form of renormalized bilinear heavy quark currents is completely
determined by the reparametrization invariant extension of the leading
order currents. This is achieved by imposing reparametrization
invariance on both the effective current operators and the
velocity-dependent short-distance coefficient functions. This way, the
velocity transfer variable $w=v\cdot v'$ is promoted into an operator
$\widehat{w}$, which in matrix elements can be replaced by a new
variable $\bar w$ that can be interpreted as being the
``short-distance'' velocity transfer of free heavy quarks, i.e., the
velocity transfer seen by hard gluons. This variable depends only on
the hadron velocities and is therefore invariant under
reparametrizations. For the vector current, the result reads
\begin{eqnarray}\label{result}
   \bar Q'\,\gamma^\mu\,Q &\,\widehat{=}\,&
    \overline{\widetilde{h}'}_{v'}\,\Big[ C_1(\widehat{w})\,\gamma^\mu
    + C_2(\widehat{w})\,{\cal{V}}^\mu\,
    + C_3(\widehat{w})\,{\cal{V}}'^{\dagger\mu} \Big]\,
    \widetilde{h}_v \phantom{ \Bigg] } \nonumber\\
   &\,\widehat{=}\,& C_1(\bar w)\,\Bigg[ J_1 + {O_1\over 2 m_Q}
    + {O_8\over 2 m_{Q'}} \Bigg]
    + C_2(\bar w)\,\Bigg[ J_2 + {O_2+2 O_4\over 2 m_Q}
    + {O_9\over 2 m_{Q'}} \Bigg] \nonumber\\
   &+& C_3(\bar w)\,\Bigg[ J_3 + {O_3\over 2 m_Q}
    + {O_{10}+2 O_{11}\over 2 m_{Q'}} \Bigg] + {\cal{O}}(1/m^2) \,,
\end{eqnarray}
with ${\cal{V}}$ and $\widetilde{h}_v$ as defined in (\ref{Vdef}) and
(\ref{htdef}), respectively. The generalization to other currents is
straightforward. The matrix elements of the effective current operators
$J_i$ and $O_j$ can be parameterized by universal functions of the
hadron velocity transfer in the standard way \cite{Falk,Luke}.
Reparametrization invariance relates the anomalous dimensions of the
dimension three and dimension four operators in (\ref{result}), and
this leads to relations between the $\mu$-dependence of the associated
universal functions.

For the vector and axial vector currents, the coefficients $C_i$ are
known to next-to-leading order in renormalization-group improved
perturbation theory, for an arbitrary ratio of the heavy quark masses.
The ingredients which go into their calculation are the one- and
two-loop anomalous dimensions of the operators $J_i$ \cite{Falk,Rady},
and the full one-loop matching between QCD and the heavy quark
effective theory \cite{QCD,AB}. In the case of different heavy quark
masses one needs in addition the anomalous dimensions and matching in
the intermediate effective theory, which governs the region $m_{Q'} <
\mu < m_Q$ \cite{Poli,JiMu,Broa}. Detailed lists of the numerical
values of $C_i$ as functions of $\bar w$ and the heavy quark masses are
compiled in Ref.~\cite{QCD}. By virtue of (\ref{result}) the currents
are now known to order $1/m$ with the same accuracy.

\acknowledgements
It is a pleasure to thank Michael Peskin for an enlightening
discussion. Financial support from the BASF Aktiengesellschaft and from
the German National Scholarship Foundation is gratefully acknowledged.

\end{document}